\documentclass[twocolumn,prl]{revtex4}
\usepackage{epsfig,amsmath,amssymb,graphics,color,calc}

\begin{document}

\newcommand{\bq}{\ensuremath{{\bf q}}}
\renewcommand{\cal}{\ensuremath{\mathcal}}
\newcommand{\bqp}{\ensuremath{{\bf q'}}}
\newcommand{\bbq}{\ensuremath{{\bf Q}}} 
\newcommand{\bp}{\ensuremath{{\bf p}}}
\newcommand{\bpp}{\ensuremath{{\bf p'}}}
\newcommand{\bk}{\ensuremath{{\bf k}}}
\newcommand{\bx}{\ensuremath{{\bf x}}}
\newcommand{\bxp}{\ensuremath{{\bf x'}}}
\newcommand{\by}{\ensuremath{{\bf y}}}
\newcommand{\byp}{\ensuremath{{\bf y'}}}
\newcommand{\bxpp}{\ensuremath{{\bf x''}}}
\newcommand{\rmd}{\ensuremath{{\rm d}}}
\newcommand{\intk}{\ensuremath{{\int \frac{d^3\bk}{(2\pi)^3}}}}
\newcommand{\intq}{\ensuremath{{\int \frac{d^3\bq}{(2\pi)^3}}}}
\newcommand{\intqp}{\ensuremath{{\int \frac{d^3\bqp}{(2\pi)^3}}}}
\newcommand{\intp}{\ensuremath{{\int \frac{d^3\bp}{(2\pi)^3}}}}
\newcommand{\intpp}{\ensuremath{{\int \frac{d^3\bpp}{(2\pi)^3}}}}
\newcommand{\intx}{\ensuremath{{\int d^3\bx}}}
\newcommand{\intxp}{\ensuremath{{\int d^3\bx'}}}
\newcommand{\intxpp}{\ensuremath{{\int d^3\bx''}}}
\newcommand{\drho}{\ensuremath{{\delta\rho}}}
\newcommand{\rhoh}{\ensuremath{{\hat{\rho}}}}
\newcommand{\fh}{\ensuremath{{\hat{f}}}}
\newcommand{\phih}{\ensuremath{{\hat{\phi}}}}
\newcommand{\thetah}{\ensuremath{{\hat{\theta}}}}
\newcommand{\etah}{\ensuremath{{\hat{\eta}}}}
\newcommand{\0}{\ensuremath{{(\bk,\omega)}}}
\newcommand{\x}{\ensuremath{{(\bx,t)}}}
\newcommand{\xp}{\ensuremath{{(\bx',t)}}}
\newcommand{\xtp}{\ensuremath{{(\bx',t')}}}
\newcommand{\xtpp}{\ensuremath{{(\bx'',t')}}}
\newcommand{\xttpp}{\ensuremath{{(\bx'',t'')}}}
\newcommand{\xtpn}{\ensuremath{{(\bx',-t')}}}
\newcommand{\xtppn}{\ensuremath{{(\bx'',-t')}}}
\newcommand{\xn}{\ensuremath{{(\bx,-t)}}}
\newcommand{\xpn}{\ensuremath{{(\bx',-t)}}}
\newcommand{\xppn}{\ensuremath{{(\bx',-t)}}}
\newcommand{\xpp}{\ensuremath{{(\bx'',t)}}}
\newcommand{\xxp}{\ensuremath{{(\bx,t;\bx',t')}}}
\newcommand{\Crr}{\ensuremath{{C_{\rho\rho}}}}

\newcommand{\Crf}{\ensuremath{{C_{\rho f}}}}
\newcommand{\Crt}{\ensuremath{{C_{\rho\theta}}}}
\newcommand{\Cff}{\ensuremath{{C_{ff}}}}
\newcommand{\Cffh}{\ensuremath{{C_{f\fh}}}}
\newcommand{\Ct}{\ensuremath{{\dot{C}}}}
\newcommand{\Ctt}{\ensuremath{{\ddot{C}}}}
\newcommand{\Crrp}{\ensuremath{{\dot{C}_{\rho\rho}}}}
\newcommand{\Crfp}{\ensuremath{{\dot{C}_{\rho f}}}}
\newcommand{\Crtp}{\ensuremath{{\dot{C}_{\rho\theta}}}}
\newcommand{\Cffp}{\ensuremath{{\dot{C}_{ff}}}}
\newcommand{\Crrpp}{\ensuremath{{\ddot{C}_{\rho\rho}}}}
\newcommand{\thetab}{\ensuremath{{\overline{\theta}}}}
\newcommand \be  {\begin{equation}}
\newcommand \bea {\begin{eqnarray} \nonumber }
\newcommand \ee  {\end{equation}}
\newcommand \eea {\end{eqnarray}}

\title{Field theories and exact stochastic equations for interacting
particle systems}
\author{Alexei Andreanov$^1$, Giulio Biroli$^{1}$, Jean-Philippe
Bouchaud$^{2,3}$, Alexandre Lef{\`e}vre$^{1}$}
\affiliation{
$^1$ Service de Physique Th{\'e}orique,
Orme des Merisiers -- CEA Saclay, 91191 Gif sur Yvette Cedex, France.}
\affiliation{
$^2$ Service de Physique de l'{\'E}tat Condens{\'e},
Orme des Merisiers -- CEA Saclay, 91191 Gif sur Yvette Cedex, France.}
\affiliation{
$^3$ Science \& Finance, Capital Fund Management, 6 Bd
Haussmann, 75009 Paris, France.}

\begin{abstract}
We present a new approach to the dynamics of interacting particles
with reaction and diffusion. Starting from the underlying discrete
stochastic jump process we derive a general field theory 
describing the dynamics of the density field, which we relate to an exact stochastic
equation on the density field. We show how our field theory maps onto the original Doi-Peliti
formalism, allowing us to clarify further the issue of the 'imaginary' Langevin noise that appears
in the context of reaction/diffusion processes. Our procedure applies to a wide class of problems
and is related to large deviation functional techniques developed recently to describe fluctuations 
of non-equilibrium systems in the hydrodynamic limit. 
\end{abstract}

\maketitle

Many problems of current interest in statistical physics involve
strongly interacting particles that
exhibit non trivial collective phenomena. One example is that of
supercooled liquids, where the dynamics 
slows down dramatically as the glass transition is approached, due to the
increasingly collective nature
of the dynamics \cite{reviewglass}. Other examples are given by systems of diffusing 
particles that branch and/or annihilate and that are subjected to an external drive; 
depending on the relative strength of these effects a variety of
non-equilibrium transitions and anomalous scaling 
behaviour can appear \cite{reviewhinrichsen-evans}. Developing theoretical techniques for such
difficult problems is of great importance, in view of the 
diversity of situations in which they can apply, and of much broader scope
than usual equilibrium statistical mechanics. 
A natural framework in this context is that of field theory which 
combined with perturbative renormalization group techniques 
have been applied successfully to a wide range of non-equilibrium phase
transitions \cite{Cardy}. Furthermore, even in non perturbative regimes, 
field theory can be very useful because it allows one to
articulate different types of 
approximations, such as the Exact Renormalisation
Group approach \cite{ERG} and Mode Coupling Theory \cite{MCT1,MCT2}. 
A field theoretical formulation
of interacting particle systems which has become somewhat standard in
recent years is provided by 
the Doi-Peliti formalism ({\sc dp}) \cite{Cardy}. Starting from a second quantization
representation of the Master equation, one obtains, 
after a rather elaborate coherent state representation, a field theory
in terms of two fields $\phi$
and $\phih$ (see below) which has been the starting point of a very
large number of studies \cite{Cardy}. 
However, besides its intrinsic complexity, the formalism
appears to be fraught with difficulties. For
example, the action of the field theory corresponds to a reasonable
looking Langevin equation for the density of 
particles, except that the noise is often complex or even pure
imaginary. This suggests that the field  $\phi$, despite 
its superficial resemblance with the density, in fact lacks direct
physical interpretation, see e.g. \cite{Tauber}. 

The aim of this Letter is to formulate a two-field theory for
interacting particles in a transparent way, starting from
a natural representation of the microscopic stochastic dynamics of the
system in discrete space. Our formalism focuses on the {\it physical} 
density field $\rho$ and, as a consequence, is related directly to 
the stochastic equations governing the evolution of $\rho$.   
The original {\sc dp} formalism is recovered performing 
a canonical Cole-Hopf transformation. 
Our method is straightforward, free of the ambiguities related to the
imaginary noise and applies to many
different situations (diffusion in an external force field, pair-wise 
interacting particles, branching/annihilation processes, hard-core
particles etc.). Furthermore it is related to recent techniques developed 
in mathematical physics (see below) \cite{Jona-Lasinio}. As an
interesting physical side product, our field theory 
in the case of liquids is found to be identical to the one obtained from
a Martin-Siggia-Rose-De Dominicis-Janssen ({\sc msrdj}) \cite{MSR} representation of the 
stochastic equation on the density field derived by Dean for Langevin
particles \cite{dean}.  This is important
since Dean's derivation contains several subtleties.
In particular, some have raised concerns about a possible hidden
coarse-graining procedure that would explain why the 
ideal gas entropy appears, quite unexpectedly, in the Langevin equation.
We end the paper with various technical comments, 
and possible applications and extensions of our formalism.  

Let us start from the simplest situation -- two sites labeled $1$ and
$2$ between which particles hop back and forth 
with a Poisson rate $W_{12}$ and $W_{21}$. The (integer) number of
particles on the two sites are $n_1$ and $n_2$. 
The variation of $n_i$ between $t$ and $t+\rmd t$ will be noted $\rmd J_i$
as is standard for {\it Poisson jump processes} \cite{Jump}; it is not a 
small quantity since it is equal to $0$ or $\pm 1$, but the probability
for it to be non zero is of order $\rmd t$. Of course,
$\rmd J_1$ and $\rmd J_2$ are strongly correlated since a particle leaving site
$1$ lands on site $2$ and vice-versa. 
More precisely, $\rmd J_1=-\rmd J_2=+1$ with probability $n_2 W_{21} \rmd t$,
$\rmd J_1=-\rmd J_2=-1$ with probability $n_1 W_{12} \rmd t$,
and $\rmd J_1=\rmd J_2=0$ otherwise. We now introduce, {\it {\`a} la}
{\sc msrdj}, the generating function for the histories
of the system: $Z(\{n,\hat n\}) = \langle \prod_t \exp [\hat
n_1(t)(\rmd J_1 - \rmd n_1) + \hat n_2(t)(\rmd J_2 - \rmd n_2)] \rangle$, 
where the averaging $\langle ... \rangle$ is over the realizations of the
Poisson jump processes. From the above rules, it
is easy to find the result in the limit $\rmd t \to 0$:
\begin{widetext}
\be
Z(\{n,\hat n\}) = \exp \int {\rmd} t \left[-\hat n_1 \partial_t
n_1-\hat n_2 \partial_t n_2 + n_1 W_{12} 
(e^{\hat n_2 - \hat n_1} -1) +  n_2 W_{21} (e^{\hat n_1 - \hat n_2} -1)
\right].
\ee
\end{widetext}
Obviously, one could add different processes, such as for example on-site 
annihilation, where $\rmd J_i = -1$ with probability 
$n_i  \mu \rmd t$, branching, where $\rmd J_i =+1$ with
probability $n_i \nu \rmd t$, two-body annihilation where $\rmd J_i=-2$ with probability 
$n_i(n_i-1)  \lambda \rmd t$, etc. With $N$ sites 
on a lattice, the total {\sc msrdj} `action' reads $S = \ln Z$:
\begin{widetext}
\begin{equation}\label{action}
S(\{n,\hat n\}) = \int {\rmd} t [-\sum_i \hat n_i \partial_t
n_i+ \sum_{\langle ij \rangle} n_i W_{ij} 
(e^{\hat n_j - \hat n_i} -1) + \sum_i n_i \left\{ \mu(e^{-\hat
n_i}-1) + \nu (e^{\hat n_i}-1) + \lambda(n_i-1)(e^{-2\hat
n_i}-1)\right\}],
\end{equation}
\end{widetext}
where $\langle ij \rangle$ means that the sum is over all couples $i,j$.
In addition, a factorized initial condition with distribution $p(n_i(0))$ can be
included by adding a contribution $\sum_i\left[-\hat n_i(0)n_i(0)+\ln
g(\hat n_i(0))\right]$, with $g(x)=\sum_q p(q)e^{qx}$. As it is customary 
in procedures {\`a} la {\sc msrdj} \cite{Cardy} 
the average value over the stochastic dynamics
of any observable $O$, a generic function of $\{n_i (t)\}$,
equals the average of $O$ over the field theory  characterized by the action (\ref{action}) and the
fields $n_i,\hat n_i$ which should be treated as {\it continuous}. This ends the derivation of the field theory that, 
as anticipated, turns out to be much more straightforward than the {\sc dp} one.  

We now want to consider the continuum limit of (\ref{action}). In this
aim we write $W_{ij} = W_0 \exp([U_i-U_j]/2T)$ 
where $U_i$ is an on-site potential (possibly time dependent), which
varies on scales much larger than the lattice
spacing $a$ and $T$ is the temperature. Restricting to nearest-neighbour hopping, we therefore
write: $W_{ij} = W_0 (1 - a {\bf e}_{ij}\cdot\nabla 
U(\vec x,t)/2T)$, where $\vec x$ is the position in space of site $i$ and
${\bf e}_{ij}$ is the unit vector pointing from $i$
to $j$. Defining the local density field $\rho(\vec x,t) = n_i/a^d$ 
and $\rhoh(\vec x,t) = \hat n_i$ and expanding to second-order in
gradients, we finally obtain (in the $a\rightarrow 0$ limit, and with $\lambda=0$ for the time being -- see below): 
\begin{widetext}
\begin{equation}
\label{eq:difflatV}
\begin{split}
S(\{\rho,\rhoh\})=&\int \rmd t \rmd \vec x \, \left[-\rhoh(\vec
x,t)\,\partial_t\rho(\vec x,t) + \rho(\vec x,t)\left( \mu
(e^{-\rhoh}-1) + \nu (e^{\rhoh}-1) \right)\right]\\ 
-&{\gamma}\int \rmd t \rmd \vec x\,\nabla\rhoh(\vec x,t)\cdot
\rho(\vec x,t)\nabla U(\vec x,t) 
+\gamma T \int \rmd t \rmd\vec x\,\left[-\nabla\rho(\vec
x,t)\cdot\nabla\rhoh(\vec x,t)
+\rho(\vec x,t)(\nabla\rhoh(\vec x,t))^2\right],
\end{split}
\end{equation}
\end{widetext}
where $\gamma \equiv W_0 a^2/2T$ is the mobility of the particles. 
Note that the above derivation is easily generalized to many other processes
and is independent of the particular form of $W_{ij}$ as long as detailed balance is verified 
(although one could also consider non-potential force fields as well). 

Before mapping this action onto the more standard  
{\sc dp} form, we want to specialize to the case where $\mu=\nu=0$,
and $U(\vec x,t)$ comes from a two-body interaction
between the particles, i.e: $U(\vec x,t) = \int \rmd\vec y \rho(\vec y,t)
V(\vec x-\vec y)$. In this case, one finds that
the above action is identical to the one obtained by using the standard
{\sc msrdj} representation for the 
following Langevin equation, derived by Dean for interacting Brownian
particles:
\be
\partial_t\rho(\vec x,t)=\nabla\cdot \left[{{\gamma}\rho(\vec x,t) \nabla
\frac{\partial F[\rho]}{\partial \rho} +\sqrt{\rho(\vec x,t)}\vec \eta(\vec
x,t)} \right],
\ee
where $\vec \eta$ is a Gaussian white noise with correlations
$\langle\eta_\alpha(\vec x,t) \eta_\beta(\vec x',t')\rangle
=2 \gamma T \delta_{\alpha\beta}\delta(\vec x-\vec x')\delta(t-t')$ and
$F[\rho]$ is the effective free-energy which has the 
naively expected shape $F[\rho] = T \int \rmd\vec{x} \rho (\vec{x}) \ln
\rho (\vec{x}) + 1/2 \int \rmd\vec{x}\rmd\vec{y} \rho (\vec{x}) V (\vec{x}-\vec{y})
\rho (\vec{y})$ [Note that Dean set the mobility $\gamma$ of the particles to unity]. This is quite remarkable,
since 
$\rho(\vec x,t)$ in Dean's equation is the exact continuum microscopic density of
the system, $\rho(\vec x,t) = \sum_i \delta(\vec x 
-\vec r_i(t))$, before any coarse-graining [$r_i(t)$ denotes the
position of particle $i$ at time $t$]; it is therefore not at all
trivial that the effective free-energy should 
have a mean-field form. The fact that our derivation, which starts on a
lattice and never uses Ito calculus, leads to the
same action confirms the validity of Dean's analysis.

Let us now consider the following Cole-Hopf change of variables, from
the above fields $\rho,\rhoh$ to new fields $\phi,\phih$
defined as:
\be
\rho(\vec x,t)=\phih(\vec x,t)\phi(\vec x,t^-); \qquad \rhoh(\vec
x,t)=\ln(\phih(\vec x,t)).\label{eq:c-h}
\ee
The Jacobian of this transformation is $1$ and thus the
measure is preserved. The reason for $t^{-}=t-\epsilon$ will be
discussed in the following. Note that this change of variable could also
be defined on the discrete lattice and it was already
considered in the literature \cite{Cardy,grass,Jan,bra}, but the precise 
connection that we establish here seems not to have been noted before.
It is straightforward to show that the above action
$S(\{\rho,\rhoh\})$ transforms into:
\begin{equation}
\begin{split}
S(\{\phi,\phih\}) =& \int \rmd t \rmd \vec x\,\left\{ \phih
\left[-\partial_t\phi+\gamma T\nabla^2\phi+{\gamma} 
\nabla(\nabla U\phi) \right]\right. \\
&\left.+  (\nu - \mu\hat\phi)\phi(1-\phih)\right\}. 
\end{split}
\end{equation}
Setting $U=0$ in the above equation, we recover {\it exactly} the
{\sc dp} action for the problem of diffusing, branching and annihilating
particle, usually derived in a rather thorny way from a second-quantization
representation of the Master equation (up to some boundary terms,
which for clarity will be discussed later). Note that
for $\mu=\nu=0$, the above action can be seen as the two-field
representation of the propagator of the non-Hermitian
Fokker-Planck operator for particles diffusing in a potential field \cite{Luck}.
In addition, in the case of two-body interaction 
between the particles, the $\nabla[\nabla U\phi]$ term becomes
$\nabla[\int \rmd\vec y \phi(\vec y,t)\phih(\vec y,t)
\nabla V(\vec y - \vec x) \phi(\vec x,t)]$. Let us finally remark that the
mapping from the action (\ref{action}) to its {\sc dp} counterpart works also on the
lattice, the continuum limit does not play any important role from this perspective. 

The above action looks very close to a {\sc msrdj} representation of the naive Langevin equation
describing the problem if one interprets $\phi$ as a density and shifts 
$\hat{\phi}\rightarrow \hat{\phi}+1$ in order to have $\langle
\hat{\phi}\rangle=0$ as in usual {\sc msrdj} (see e.g. \cite{Cardy}). 
However this interpretation is problematic since, as mentioned in the
introduction, the noise term is unphysical \cite{Tauber}. For example, 
in the case of diffusing particles with pair-wise interaction ($\mu=\nu=0$) 
the ``noise'' term has a correlator $\langle\eta(\vec x,t)\eta(\vec
x',t')\rangle= 2\nabla_x [\phi(\vec x)(\nabla_x V)(\vec x-\vec x')
\phi(\vec x')]$. For a uniform density $\phi$ and $V(x)=V_0\exp(-x^2)$, the variance of the noise 
is a negative definite operator meaning that the noise has an imaginary part! On the other hand, we
clearly see that this difficulty disappears when 
one uses the fields $\rho,\rhoh$, which encode a very well
defined underlying Langevin equation, albeit with non Gaussian, Poisson jump terms corresponding 
to particle creation and annihilation.

Let us now focus on some technical but important subtleties of the transformation
(\ref{eq:c-h}). Consider pair-annihilation ($A+A\rightarrow 0$) with rate
$\lambda$ on a single site. The corresponding contribution in the {\sc msrdj}
and {\sc dp} actions are $\lambda \rho(\rho-1)\left(e^{-2\rhoh}-1\right)$
and $\lambda\phi^2(1-{\hat\phi}^2)$, which do not transform exactly into
one another under (\ref{eq:c-h}). 
The underlying reason is that
the two field theories correspond to different time discretizations. This 
makes a difference when the action contains non linear terms evaluated at the same
position in time and space, because the response function $\langle \phi(\vec x, t)
\hat \phi(\vec x, t')\rangle$ is discontinuous when $t=t'$. This can be traced back to
the fact that the fields $\phi$ and $\hat\phi$ are the coherent state
representations of creation/annihilation operators of the theory, $a$ and $a^+$, 
which satisfy $[a,a^+]=1$. In the {\sc dp} formalism the action is obtained after normal
ordering \cite{Cardy} and as a consequence, $\hat \phi(\vec x, t) \phi(\vec
x, t) \hat \phi(\vec x, t) \phi(\vec x, t)$ is in fact the continuous time limit 
of $\hat \phi(\vec x, t) \hat \phi(\vec x, t-\epsilon) \phi(\vec x, t-2\epsilon) \phi(\vec
x,t-3\epsilon)$. Instead, the field theory obtained from (\ref{action}) through
the transformation (\ref{eq:c-h}) has no normal ordering and therefore the term 
$\lambda \rho^2(\vec x, t)$ is the continuous time limit of $\lambda \hat \phi(\vec x, t) 
\phi(\vec x, t-\epsilon) \hat \phi(\vec x, t-2\epsilon) \phi(\vec
x,t-3\epsilon)$. Thus, to transform our field theory into {\sc dp}, one
has to use the transformation (\ref{eq:c-h}) and in addition to take care 
of the time discretization which amounts to do normal ordering, and to recover the 
$-1$ missing in the above example. 
An alternative way to make the exact connection between the two field theory is
through the operator formalism. After having expressed the Master equation
in terms of the operators $a,a^+$ one does a canonical transformation 
$a=e^{-\rho^+}\rho$, $a^+=e^{\rho^+}$ such that $[\rho,\rho^+]=1$.
The operators $\rho,\rho^+$, originally introduced in \cite{MSR}, 
lead to a different representation of the Master equation. This then leads exactly, using again a coherent state 
representation, to the field theory that we derived above directly from underlying the stochastic process, including
the correct boundary terms.

Another case where our strategy applies is interacting particles evolving 
under Newtonian dynamics, for which Doi also derived a field theory \cite{Doi}. Our
previous starting point consisted in deriving exact stochastic equations for the density 
field and then applying the {\sc msrdj} procedure to get the field theory. For Hamiltonian dynamics, 
one can derive an exact deterministic equation on 
the density in position $\vec r$ and momentum $\vec p$ space, $\rho(\vec x,\vec
p;t)=\sum_{i}\delta (\vec{r}-\vec{r}_{i})\delta (\vec{p}-\vec{p}_{i})$:
\begin{equation}
\begin{split}
\partial_t\rho(\vec x,\vec p;t)&=-\frac{1}{m}\vec p\cdot\partial_{\vec
    x}\rho(\vec x,\vec p;t)\\ 
+\int d \vec x' d \vec p'\rho(\vec x',\vec p';t)&\partial_{\vec
  x}V(\vec x-\vec x')\cdot\partial_{\vec p}\rho(\vec x,\vec p;t).
\end{split}
\end{equation}
In this case only the initial conditions are stochastic. The {\sc msrdj} field theory
corresponding to this equation maps exactly onto Doi's  
using the transformation, akin to (\ref{eq:c-h}),
$\phi(\vec x,\vec p;t)=e^{-\rhoh(\vec x,\vec p;t)}
\rho(\vec x,\vec p;t)$ and $\hat\phi(\vec x,\vec p;t)=e^{\rhoh(\vec
  x,\vec p;t)}$ (note that $\rhoh(\vec x,\vec p;t)$ is the {\sc msrdj} field
conjugate to $\rho(\vec x,\vec p;t)$).

Finally, our formalism in the continuum limit bridges the gap 
between the purely microscopic Fokker-Planck evolution operator and the
hydrodynamics description studied recently in the mathematical physics
literature. A method to handle both the hydrodynamic limit and large
and rare fluctuations around it, has been developed in \cite{Jona-Lasinio}. The starting point of 
this work is very close to the generating functional we used. However, in order to focus on hydrodynamics
length-scales and timescales, as done in \cite{Jona-Lasinio}, one has to consider a conjugated field 
$\hat n_i=\hat \rho(x/L)$ that is constrained to varying only on length-scales of the order of the
linear system size $L$. In this case the ``action'' or functional (\ref{action})
becomes a function of the hydrodynamic density field that represents the
average density inside very large boxes, and can be related to the rate
functional introduced and studied in \cite{Jona-Lasinio}. 
On the other hand, the stochastic equations corresponding to our 
continuum limit are valid on a scale $\ell$ much larger than the lattice spacing $a$ but
much smaller than the system size $L$. As a consequence these allow one to 
tackle, with field theoretical techniques, dynamic phase transitions where
the physically relevant length-scales are much larger than $a$ but not necessarily much
larger than the (diverging) correlation length $\xi$. The hydrodynamic limit of \cite{Jona-Lasinio} instead
corresponds to length-scales much larger than $\xi$.

In conclusion our procedure allows one to derive rather straightforwardly 
a field theory different from, but dual of, the  Doi-Peliti formalism. This
could be useful in cases where this by now standard framework does not work,
see e.g. \cite{JanssenTauber}. It certainly avoids, contrarily to {\sc dp}, very cumbersome
computations due to normal ordering in cases in which the rates are
complicated functions of the local density. Another advantage of our approach is that 
it is, almost by construction, directly related to stochastic equations on the density field. 
The representation in terms of a stochastic equation, especially after
having taken the continuum limit, is particularly appealing. It can be helpful 
for numerical investigations \cite{Chate} since it might be more efficient
to integrate numerically than to simulate the original lattice model.
Furthermore, stochastic equations are very useful to encode and study universality 
classes as it has been understood in the case of critical slowing down
\cite{Halperin} and non-equilibrium phenomena such as surface growth \cite{KPZ}. 
Some applications of the results of this letter are underway, for example 
the study of the condensation phase transition in the zero range process
\cite{bgl}. From a more general perspective our results make clear that
exact stochastic equations can be always obtained, thereby avoiding
phenomenological guesses which, especially for off-equilibrium cases, can be 
very tricky \cite{Munoz}.

{\em Acknowledgments - } We thank I. Dornic, C. Godr\`eche, F. Van Wijland for helpful
and interesting discussions. A substantial part of this work has been done at
the Isaac Newton Institute for Mathematical Sciences during the program 
``Principles of the Dynamics of Non-Equilibrium Systems''. AA and GB are partially
supported by the European Community's Human Potential Program
contracts HPRN-CT-2002-00307 (DYGLAGEMEM).

\end{document}